# Blame is easier than praise: Measuring off-ball defensive performance in football

Jonas Bischofberger*[1,2,3]✉, ORCID: 0000-0002-0878-0641, Runqing Ma*[1,2]✉, ORCID: 0009-0007-4250-4687, Pascal Bauer[4,5], ORCID: 0000-0001-8613-6635, Kilian Arnsmeyer[5], Arnold Baca[1], ORCID: 0000-0002-1704-0290

* Equal contribution

[1] Centre for Sport Science and University Sports, University of Vienna, Vienna, Austria
[2] Vienna Doctoral School of Pharmaceutical, Nutritional and Sport Sciences (VDS-PhaNuSpo), University of Vienna, Vienna, Austria
[3] VfB Stuttgart 1893 AG, Stuttgart, Germany
[4] Universität des Saarlandes, Saarbrücken, Germany
[5] Deutscher Fußball-Bund e.V. (DFB), Frankfurt, Germany

✉ jonas.bischofberger@univie.ac.at

✉ runqing.ma@univie.ac.at

# Abstract

The defensive performance of football players is commonly measured through a limited number of actions like tackles and interceptions while their continuous impact through positional behaviour has hardly been studied before. We formulate this problem as an attribution over multi-agent spatiotemporal trajectories without player-level ground truth labels, where event-level changes of expected threat are distributed among individuals. We propose a framework that performs this attribution using player involvement scores calculated from defensive pressure areas (DPAs). By computing role-conditioned baselines within automatically detected team structures, we can determine each defender's expected responsibility for threat created through arbitrary passes.

The validity and robustness of this approach are evaluated on a uniquely extensive cross-gender and cross-competition data set, including positional and event data from 64 matches of the men's World Cup, 116 matches of the women's German Bundesliga and 336 matches of the men's German 3. Liga. In the absence of a ground truth, we propose an evaluation protocol that combines multiple relatively weak proxies into robust summary scores. We find a validity score that is improved by around 1 standard deviation compared to the best action-based metric and demonstrate that many popular measures show limited validity. The "blame" for conceding high-value actions shows especially strong correlations with external ratings and market values, making it the first published metric in football to reliably measure positioning errors. All code underlying this work is publicly available to support reproducibility and further research.

# 1 Introduction

"Offence wins games, defence wins championships" (Robst et al. 2011), this famous sports proverb has been passed down through generations of sports coaches and athletes. Although being of at least equal importance, defending is a far less easily measurable aspect of team sports compared to attacking performance (Villa, Lozano 2019; Ruan et al. 2022). The simple counting of interceptions, blocks and tackles is not sufficient, as these metrics are hard to define objectively and biased towards active participation, neglecting subtle but decisive contributions through positioning and anticipation which help their team maintain stability or support other players in winning the ball (Trainor 2014; Fernandez-Navarro et al. 2016; Toda et al. 2022). A memorable example is central defender Virgil van Dijk who has been universally praised for his defensive abilities which earned him the second place at the Ballon d'Or 2019, the most prestigious individual trophy in men's football. Yet, his number of tackles, interceptions and blocks during the corresponding domestic season merely place him in the 11th, 19th and 11th percentile respectively (FBref 2025).

The advent of tracking data has opened the door for advanced defensive performance metrics that capture positional and spatial information. One strand of research in this area is focused on analysing "pressing", the organized attempt of a team to exert pressure on the opponent and disrupt their attacking play (Andrienko et al. 2017; Bauer, Anzer 2021; Merckx et al. 2021; Lee et al. 2025). These studies typically use ball-wins as the outcome variable and try to relate them to team-level behaviours (Forcher et al. 2024) while player-level analysis revolves around contributions through pressuring opponents.

The team-level analysis of defensive organization (Bauer et al. 2023; Umemoto, Fujii 2023; Eigenrauch et al. 2024) is more widespread than player-level attribution of defensive contributions. Some of them can spot weaknesses spatially (Stöckl et al. 2021; Ogawa et al. 2025) but lack the ability to distribute responsibility directly among individual players. Notably, Le et al. (2017) perform some limited player attribution by comparing player trajectories with "ghosted" trajectories predicted by a machine learning model. However, they found no straightforward correlation of being "out of position" with defensive success, indicating that their model may identify both moments of "unconventional" as well as "bad" defending. Wu and Swartz (2023) present a related approach where player velocities are compared with a “ghosted” velocity vector that indicates whether a player reacts to a given situation more rapidly than expected, providing a specific measure of anticipation skill. Llana et al. (2020) developed a classifier to detect passes behind the back and attribute the blame for such passes using normalized space control maps.

Current studies, both at the team and player level, still miss an important aspect of defensive evaluation: how the outcome of a defensive action affects the opponent’s attacking play. Although some studies consider such outcomes (Merhej et al. 2021; Rahimian, Toka 2024), they are typically anchored on recorded defensive actions (e.g., pressures, tackles, interceptions) as the starting point. This focus overlooks situations where no action is recorded, for example when a defender is out of position, fails to provide cover, or leaves a passing lane open, even though these may directly enable valuable attacking opportunities. As a result, negative or missing defensive contributions are largely absent from the evaluation.

This gap motivates a shift in perspective: instead of measuring defence only from the defender’s perspective, this study evaluates it through the lens of the attacking team’s passing outcomes, assigning blame and credit for opponent actions among defenders to account for active and passive individual contributions. To this end, efficient geometric rules are used to measure the degree of involvement in defending passes and derive defensive responsibility based on the long-running average involvement depending on the tactical role of a player. The metrics proposed in this study are validated extensively across three different data sets, covering both elite and lower-level professional football as well as women’s and men’s competitions, constituting the first comprehensive evaluation of metrics related to individual defensive positioning performance in the literature. To the best of our knowledge this is also the first study using a near-full season of women’s event and positional data. With this thorough evaluation and the publication of the code of our model (Bischofberger 2025), we hope to build the foundation for novel developments in this under-studied area of sports analytics.

Section 2 details the data set used in this study. Section 3 lays out the modelling framework behind involvement- and responsibility-based metrics. Section 4 describes the evaluation routine while Section 5 shows its results. Section 6 provides a detailed discussion of the results, followed by Section 7 which discusses the limitations of the presented work. Section 8 briefly concludes the work.

## 2 Data

For model construction and validation, we use an extensive multi-league and cross-gender spatiotemporal football dataset, as presented in Table 1. The data set comprises the Men's FIFA World Cup 2022 and the German domestic leagues Frauen-Bundesliga and 3. Liga during the 2023/24 season. The Frauen-Bundesliga is known to be one of the world's strongest domestic women's football competitions, representing an elite performance level as well as the men's World Cup. The 3. Liga represents the men's third tier of the German league system with a significantly lower level of competition. It is a professional, single-track competition that is known for a more combative playing style compared to top-level competitions.

Table 1. Overview of the multi-source dataset

| Competition | Season | Matches | Total passes | Total frames | Data provider | |
|---|---|---|---|---|---|---|
| | | | | | Tracking data | Event data |
| Men's World Cup | 2022 | 64 | 29,931 | 11,849,751 | PFF FC (30 fps) | PFF FC |
| Google Pixel Frauen-Bundesliga | 2023/2024 | 116 | 100,128 | 16,784,449 | STS/Track 160 (25 fps) | STS |
| Men's 3. Liga | 2023/2024 | 336 | 283,703 | 49,286,358 | | |

The tracking data consist of high-frequency optical spatiotemporal recordings at 25–30 frames per second (fps). For each match, the locations (X and Y coordinates) of all players and the ball are continuously captured, generating over 130,000 frames per match. Tracking data of the leagues has been captured by Sportec Solutions (STS) and Track160 using optical tracking. Due to quality issues in individual matches, the season 2023/24 is not completely covered: 16 matches (12.1%) from the Frauen-Bundesliga and 44 matches from the 3. Liga (11.6%) are absent from the provided data set. The World Cup data originates from PFF FC (FC 2025), a commercial data provider that generates tracking data from existing broadcast video.

The player positions are enriched by event data which comprise more than 400,000 passing events in total, including the start and end locations of the pass, the identities of the passer and receiver, and the pass outcome (e.g., successful, intercepted). These events were manually recorded by trained operators from the data providers STS and PFF FC. The manual nature of event data collection together with the lack of a previous validation study of the employed tracking systems should be kept in mind as a potential limitation of this data set.

The German domestic data is provided in the Common Data Format (CDF) (Anzer et al. 2025), a unified format for football match data that simplifies the process of performing analyses across data sets. The proprietary format of the World Cup data was converted into the CDF to be processed in unison. To align the timestamps of the tracking and event data of the German competitions, we use the synchronization supplied in the CDF (Anzer, Bauer 2021; Anzer et al. 2025). For the World Cup data, we use the provider's native synchronization. An example frame of synchronized data is shown in Figure 1.

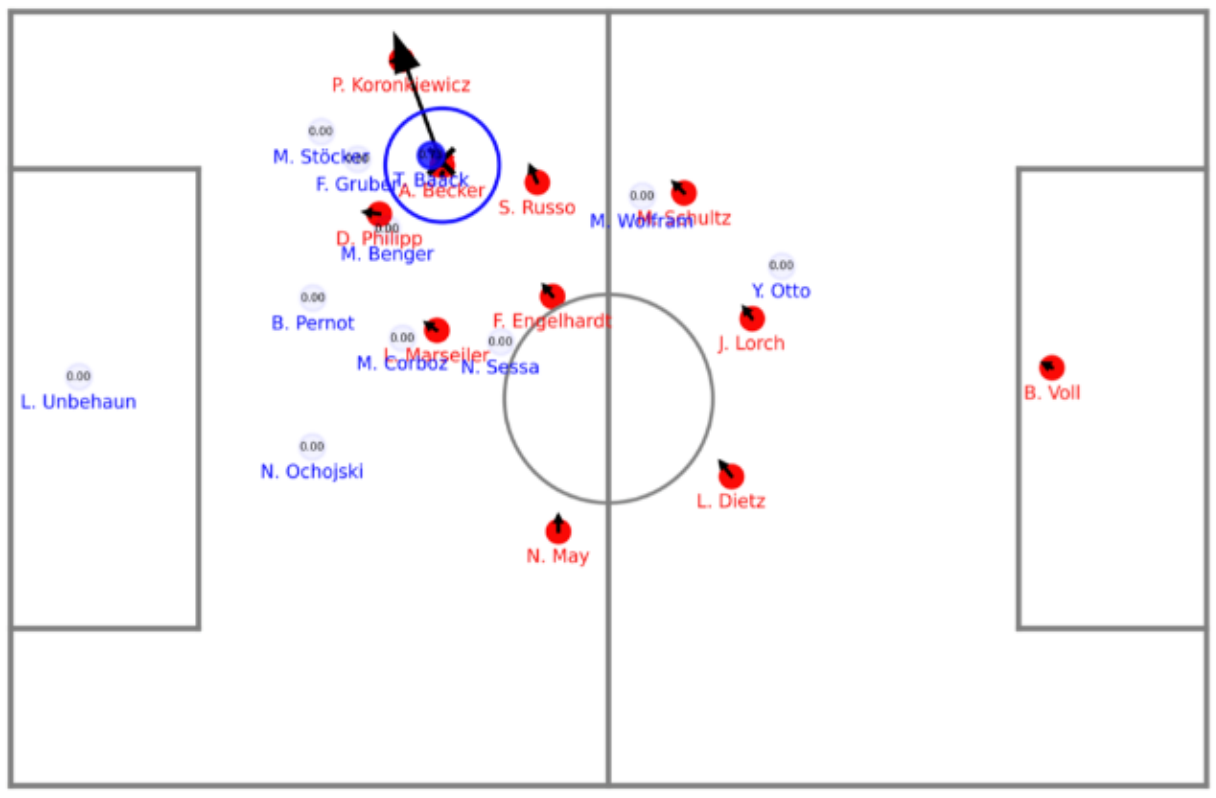

Figure 1. Example of synchronized data during a pass, with the position of all players on the field, red dots for attacking players, blue dots for defending players, a black cross for the ball and a directed arrow for the pass trajectory.

## 3 Model specification

The model follows several consecutive steps: During pre-processing, all features required to model defensive attribution are calculated: Pass value, measured through Expected Threat (Singh 2018), the expected receiver of failed passes, and the role of each player based on dynamically calculated team formations. Second, the degree of *involvement* of every defender with respect to every pass in the match is determined based on their spatial proximity and interception actions. Third, the involvement values are averaged for every triplet consisting of: (a) passer role, (b) (intended) receiver role and (c) defender role to obtain a measure of expected involvement, or *responsibility*, and applied back to each individual pass. Finally, involvement- and responsibility-based metrics are aggregated and normalized to obtain final player-role-level KPIs. The entire process is shown in Figure 2.

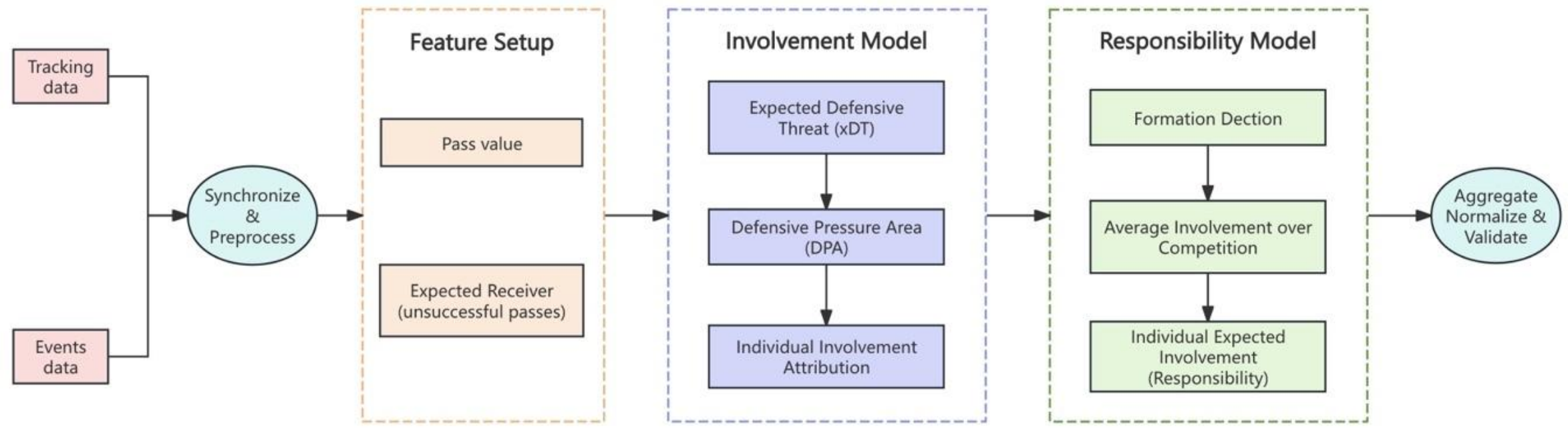

Figure 2. Overview of the modelling pipeline.

## 3.1 Defensive involvement

The evaluation of defensive performance in this study begins with the attacking team's passing actions. Each pass receives a value that reflects its impact on the attacking progression. From the defensive perspective, the same pass implies an opposite contribution: if the pass increases the attacking threat, it represents a defensive failure, whereas if it decreases the attacking threat, it reflects defensive success.

To attribute this defensive value to individual players, the model incorporates the spatial context of defending by defining *defensive pressure areas* (DPAs) during the moment of the pass depending on the outcome of the pass. Defenders located within this area or performing an interception are considered involved in the action and thus either effectively contribute to a decrease in attacking threat or fail to intervene in a pass that increases attacking threat.

### 3.1.1 Expected Defensive Threat (xDT)

Expected Threat (xT) (Singh 2018) is chosen to measure the value of passes, as it is known to be robust, which is essential for being used as a component of a more complex modelling routine (Van Roy et al. 2020). xT represents the probability of scoring a goal within the next few actions based on a fixed spatial grid. The model estimates scoring and transition likelihoods within a Markov chain from historical events to generate its final value estimate.

In this study, the pitch is divided into $16 \times 12$ grid cells following Singh (2018), a resolution that preserves sufficient passing information while avoiding excessive sparsity or heterogeneity. The xT model is fitted independently to the Hudl StatsBomb open data set (Statsbomb 2024) of the men's World Cup 2022 to reduce bias in xT estimates. Figure 3 shows the xT distribution across the pitch.

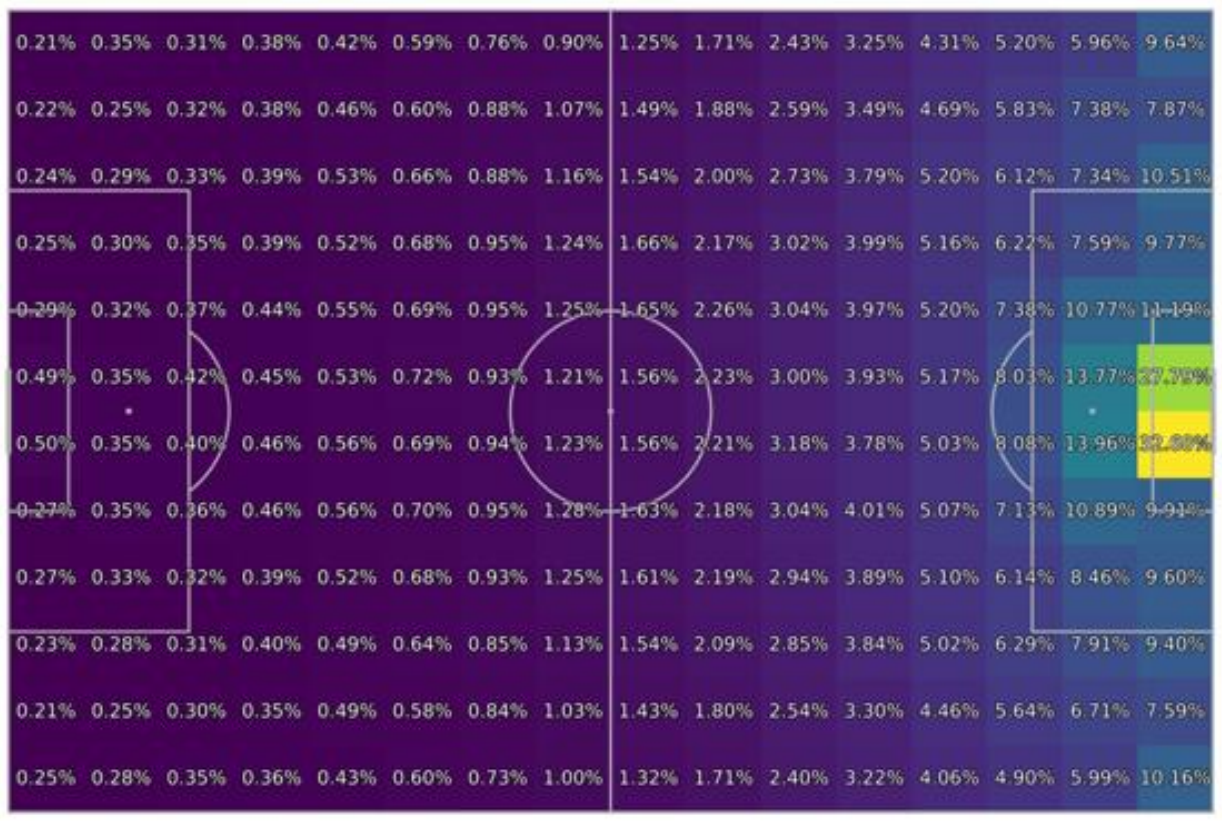

Figure 3. The xT plot of the World Cup 2022, for each zone.

The value of a pass is calculated as the difference in xT between its start and end location:

$$\Delta xT = xT_{end} - xT_{start} \tag{1}$$

A positive value means that the pass increased the team's scoring probability, whereas a negative value indicates that the pass reduced the team's scoring probability.

While passing events originate from the attacking side, the defensive perspective is evaluated as **Expected Defensive Threat ($xDT$)**. For successful passes that are completed and received by an attacking teammate, the corresponding $xDT$ value is the negative value of $\Delta xT$. For instance, a highly valuable pass for the attacking team with $\Delta xT = +0.2$ corresponds to a negative defensive value $xDT = -0.2$ for the defending side, indicating a failure to prevent a dangerous pass. For unsuccessful passes, $xT_{end}$ can be considered zero as the team loses possession, which accordingly results in a positive value for $xDT$ that is equal to the offensive value of the starting location of the pass.

$$xDT = \begin{cases} -\Delta xT, & if\ the\ pass\ is\ successful \\ xT_{start}, & if\ the\ pass\ is\ unsuccessful \end{cases} \tag{2}$$

### 3.1.2 Defensive Pressure Areas (DPAs)

In football, defensive responsibility can rarely be pinpointed to a single individual but is usually shared among multiple defenders. An initial attempt to distribute responsibility can be based on spatial proximity to the pass: A defending player who is close to the action receives credit or blame for the value that this action generates. To identify which defenders are involved in a certain pass event, we propose to define a Defensive Pressure Area (DPA). The DPAs define a spatial range at the moment of the pass within which players are considered capable of preventing the pass or exerting pressure. By establishing this area using detailed positional data, defensive pressure can be assigned to the involved defenders.

Based on the outcome and the value of the pass, four types of assignments are distinguished. The definitions and corresponding visualization are shown below and in Figure 4:

1. **Successful passes with negative $xDT$ :** Defensive involvement is considered around both the passer, the receiver and along the pass lane, represented by circles of radius r = 5 m and a rectangular zone connecting them. This area accounts for a failure to prevent both the pass and the reception and a failure to intercept along the pass trajectory.
2. **Successful passes with positive $xDT$:** Defensive pressure is considered only around the passer (circle of radius r = 5 m). These are typically backward passes, where defenders force the passer to play the ball backwards, thus making the defenders around the passer most relevant.
3. **Unsuccessful passes going out of play:** Defensive pressure is considered only around the passer (circle of radius r = 5 m).
4. **Unsuccessful passes intercepted:** In this case, no DPA is defined, as the defensive value is directly attributed to the specific defender who intercepts the ball.

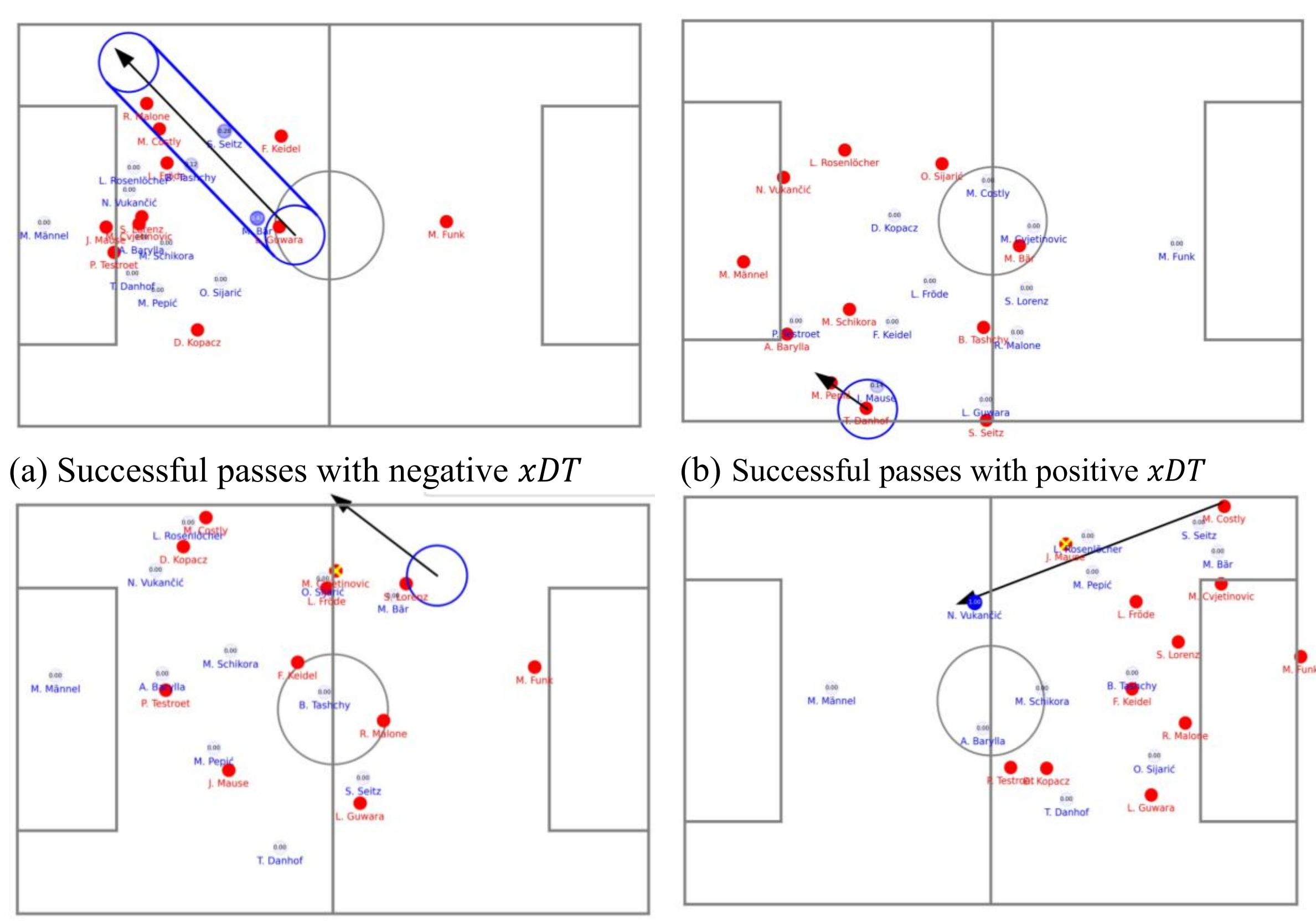

(a) Successful passes with negative $xDT$ (b) Successful passes with positive $xDT$

(c) Unsuccessful passes going out of play (d) Unsuccessful passes intercepted

Figure 4. Four types of DPAs, depending on the value and outcome of the pass. The yellow cross indicates the expected receiver of an unsuccessful pass.

The radius of 5 meters is chosen to represent a rough estimate of the area of influence of a defender. A lower radius might risk missing actual pressure and involvement while a higher radius might lead to a weaker attribution between players and stronger intra-team correlations in the final metrics.

Since there is no ground truth (Davis et al. 2024) for defensive involvement or the individual attribution of offensive value, the specific areas are handcrafted rather than determined empirically. Data-driven modelling or parameter optimization with respect to unreliable benchmarks would risk overfitting and reproducing biases present in the selected target variable. Reachable areas based on motion models (Renkin et al. 2022) were also not used, as physically reachable areas may overestimate the involvement of players who are additionally restricted by team tactics. The relation between potential motion, pressure, and possible interception would introduce additional modelling complexity. Instead, a simpler representation of spatial proximity based on reasonable geometric assumptions was adopted to maintain model simplicity, robustness, and computational efficiency.

### 3.1.3 Individual defensive attribution

For defenders located within each DPAs, involvement is distributed according to their relative proximity to the pass. The closer a defender is to the passer or the pass trajectory, the greater the defensive pressure they are assumed to exert. For pass $p$ with value $xDT_p$, the *raw proximity* $R_{p,i}$ and the *valued proximity* $V_{p,i}$ for player $i$ are defined as:

$$R_{p,i} = \frac{(r - d_i)}{r} \tag{3}$$

$$V_{p,i} = xDT_p \times R_{p,i} \tag{4}$$

where $d_i$ is the Euclidean distance from player $i$ to the edge of the DPA and $r$ is the radius of the circles shown in Figure 4.

The values of raw proximity $R_{p,i}$ and valued proximity $V_{p,i}$ can be interpreted depending on their real football meaning: $R_{p,i}$ and $V_{p,i}$ for passes with positive xDT (i.e. negative xT) reflect a defender's successful participation in suppressing attacking threat and are therefore referred to as raw ($R_{p,i}$) and valued ($V_{p,i}$) *contribution*. $R_{p,i}$ and $V_{p,i}$ for passes with negative xDT (i.e. positive xT) reflect a defensive failure, indicating that the defender was involved in but did not prevent an increase in attacking threat; this is referred to as *fault.* To account for overall participation regardless of the outcome, we also define *involvement* as the sum of the absolute values of contribution and fault. Details and descriptions of these three types of metrics are shown in Table 2 below. They allow each passing event to be decomposed into player-level defensive attributions, distinguishing between effective contributions, defensive failures, and overall involvement.

Table 2. Involvement-based defensive performance metrics for individual player $i$

| Metric | Symbol | Equation | Range | Interpretation |
|---|---|---|---|---|
| Raw contribution | $C_r$ | $C_r = \sum_{p \; if \; xDT_p \geq 0} R_{p,i}$ | $[0, \infty)$ | Intercepting and pressuring the opponent into low-value passes. |
| Valued contribution | $C_v$ | $C_v = \sum_{p \; if \; xDT_p \geq 0} V_{p,i}$ | $[0, \infty)$ | |
| Raw fault | $F_r$ | $F_r = \sum_{p \; if \; xDT_p < 0} R_{p,i}$ | $[0, \infty)$ | Allowing dangerous passes, "getting outplayed" |
| Valued fault | $F_v$ | $F_v = \sum_{p \; if \; xDT_p < 0} \lvert V_{p,i} \rvert$ | $[0, \infty)$ | |
| Raw involvement | $I_r$ | $I_r = C_r + F_r$ | $[0, \infty)$ | Defensive activity |
| Valued involvement | $I_v$ | $I_v = C_v - F_v$ | $(-\infty, \infty)$ | Overall performance |

## 3.2 Formation detection & responsibility

A limitation of measuring defensive performance through spatial proximity is that this approach does not take into account whether a player is caught out of position. For example, a defender might cause high-xT passes by leaving their assigned role vacated but would not receive any blame in terms of involvement as he or she was located far away from those passes.

Our solution is to model *defensive responsibility* for an action as the expected involvement based on specific features that determine which player "should" defend a pass. Since, in practice, defensive responsibility is usually tied to the role of the player within a team's tactical setup, we model responsibility as the average involvement for each triplet of (1) the role of the passer, (2) the role of the receiver and (3) the role of the defender. With larger amounts of data available per competition, more variables like the position of the ball or the identity of the team could be considered. The choice of variables in this work reflects a trade-off between robustness and informativeness of responsibility estimates.

In order to assign roles, the formation of both teams is identified using a template matching algorithm. Formations are determined separately for each phase of play in which one team maintains continuous possession of the ball. Both teams' X and Y positions are averaged, z-score standardized for each ball-in-play phase, and matched against a catalog of seven formations with 11 players, three formations with 10 players and one formation with 9 players. Players are categorized into 20 different roles within the following 6 role groups: Centre Back (CB), Full Back (FB), Central Midfielder (CM), Central Attacking Midfielder (CAM), Attacking Winger (Winger), and Central Forward (CF). The resulting roles and formations are shown in Figure 5. The choice of available templates and roles is made to balance the variety of formations with the number of resulting examples per role triplet for robust estimates of responsibility. The modelling approach does not depend on any particular set of roles or formations.

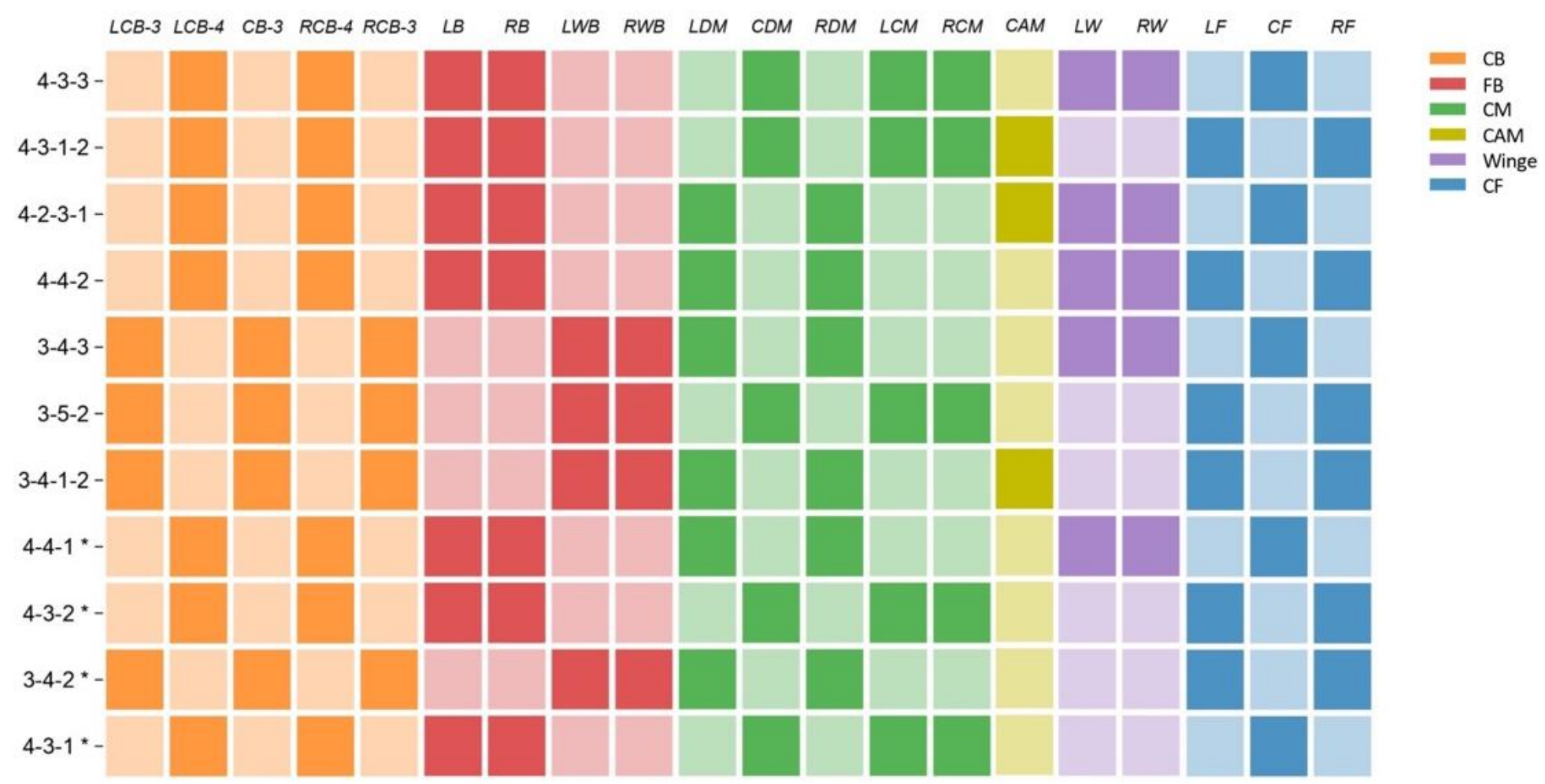

Figure 5. Presence of roles within each considered formation. Formations marked with * consist of fewer than 11 players.

A matching distance between the real player positions and each template is calculated by solving a linear sum assignment with the Euclidean distance as the cost function. The assignment costs are then smoothed using a Gaussian kernel with a standard deviation of 7.5 minutes to avoid spurious formation transitions due to temporary positional adjustments. The formation is then determined per phase according to the lowest smoothed cost value.

Based on the roles that result from the formation detection, *responsibility* is defined as the average involvement over the roles of the passer, receiver, and defender, separately for each competition. One example of responsibility values for passes from a left back (LB) to a left winger (LW) is given in Table 3. From this type of pass, the most involved roles are right winger (RW), right wing-back (RWB), and right central midfielder (RCM). A player who is currently assigned to a certain defensive role receives his or her responsibility for a pass according to the model, irrespective of his or her actual location on the pitch.

Table 3. Example responsibility values from the FIFA Men's World Cup 2022. Responsibility values represent the average involvement for each combination of roles over the entire competition.

| Passer role | Receiver role | Defender role | Responsibility |
|---|---|---|---|
| LB | LW | RW | 0.196 |
| LB | LW | RWB | 0.151 |
| LB | LW | RCM | 0.130 |
| LB | LW | RS | 0.109 |
| LB | LW | RDM | 0.085 |
| … | … | … | … |
| LB | LW | CB-3 | 0.000 |

To obtain the missing receiver role for unsuccessful passes, the expected receiver for these passes is estimated using the method proposed by Power et al. (2017). It finds the most likely receiver out of all possible receivers as the player $i$ with the lowest distance to the end point of the pass and the lowest angular deviation from the line of pass according to the following Equation (5).

$$Expected\ Receiver = \arg min_i\left(\frac{Distance_i}{min_j\ Distance_j} \times \frac{Angle_i}{min_j\ Angle_j}\right) \quad (5)$$

Here, $i$ and $j$ index candidate receivers, and min denotes the minimum value across all potential receivers.

The raw responsibility values exemplified in Table 3 can be used instead of the raw proximity values $R_{p,i}$ to obtain a corresponding responsibility-based metric for any involvement-based metric. Analogously to Table 2, we call the responsibility-based metrics *responsibility, fault responsibility*, and *contribution responsibility*.

### 3.4 Metrics aggregation

Since all metrics are defined on the level of individual events, a fine-grained aggregation of player metrics can be performed at the levels of match, player, and player role. In this way, changing responsibilities due to positional and team-tactical switches within a match are considered.

After summing up position-specific match-level totals, two aggregation methods are used to arrive at final player-position-level metrics: *per 90 minutes* and *per pass*. Aggregating per 90 minutes means taking the total of a metric and dividing it by the total number of played minutes times 90. While *per 90 minutes* is the typical way of aggregating performance data in football analytics, an additional aggregation *per pass* of the opponent team is also calculated to normalize pass-related defensive metrics by the number of opportunities to defend (Antonio 2013).

Preliminary experiments show that the validity of fault-related metrics is higher when aggregated per 90 minutes whereas contribution-based metrics perform better when aggregated per pass. To obtain promising combined metrics, an equation is proposed to fuse these aggregates into a combined involvement or responsibility as follows:

$$I_{fused} = \frac{C_{total}}{passes\ faced} \times 250 - \frac{F_{total}}{minutes\ played} \times 90 \quad (6)$$

Where $I_{fused}$ is the fused metric, $C_{total}$ is the total contribution and $F_{total}$ is the total fault. The value 90 corresponds to the typical number of minutes within a match and 250 corresponds to a rough estimate of the typical number of passes faced in a match combined with an additional weighting factor between fault and contribution. Equation (6) is calculated for the following four metrics: raw involvement, valued involvement, raw responsibility, and valued responsibility, each using their corresponding fault and contribution metric.

# 4 Model Validation

Involvement- and responsibility-based metrics are benchmarked against current state-of-art metrics that are commonly used to measure defensive contributions: In particular, tackles won, tackles won ratio, and interceptions are considered, where the number of interceptions is of particular interest as it is the only widespread metric that also captures the quality of defensive positioning. These metrics are readily available in all present datasets and widely used in the football analytics sphere (Freitas et al. 2023).

To arrive at a formal estimate on whether our metrics provide a benefit over existing ones, their validity and robustness are compared across three different data sets that span genders and competition levels.

## 4.1 Validity

Estimating the validity of defensive player metrics is difficult as there is no gold standard available. To get an estimate of the plausibility of our metrics, we examine the correlation coefficient $r_{MV}$ of our metrics with market values from "transfermarkt.de" and "soccerdonna.de" (Donna 2025; Markt 2025) as well as the correlation coefficient $r_{FIFA}$ with "defensive awareness" ratings from the video game series FIFA (Futbin 2025).

Transfermarkt.de and soccerdonna.de provide crowd-sourced estimates of a player's value on the transfer market which give a rough estimate of a player's economic value which in turn is intimately tied to their overall performance level. The values are sourced from December 2023. FIFA ratings are compiled by an extensive data collection system supplied by a worldwide network of scouts and databases and are broken down by specific skills such as "defensive awareness". The ratings are sourced from the 2024 and 2022 editions, corresponding to the respective seasons of our tracking and event data.

Both sources provide external ratings with large qualitative contributions that provide an external and objective benchmark for our metrics. Since defensive positioning skill is subtle, strong confounding factors are expected in these benchmarks. For example, a defender who is generally considered elite in terms of other defending skills such as tackles or athleticism might also be rated as a player with high positioning skill due to the Halo effect (Nisbett, Wilson 1977). Therefore, to strengthen the signal in the external ratings, the correlation analysis only includes central defenders, for whom defensive positioning is most relevant to determine overall performance. Player-role combinations below 300 minutes (3. Liga) and 150 minutes (Frauen-Bundesliga, World Cup) are also excluded from the analysis to reduce the sampling error of individual performance estimates.

## 4.2 Robustness

Additional to showing whether the metrics accurately measure defensive positioning skill based on external ratings, internal properties of these metrics are also measured to represent robustness and discriminatory ability (Franks et al. 2016). To measure these properties, three measures are used: (a) the intraclass correlation coefficient $ICC_{Match}$ on the match level to determine the degree to which match-by-match performance is repeatable and has discriminatory power, (b) the Pearson correlation $r_{Repeat}$ of the KPI in the first half of the season with the KPI in the second half of the season to see if performance is typically repeated over longer periods of time and (c) the season-level intraclass correlation coefficient $ICC_{Season}$, obtained by bootstrap sampling and aggregating all match performances to measure the robustness of season aggregates.

For the FIFA Men's World Cup data, $r_{Repeat}$ is excluded as the number of matches per team becomes too low when cut in half. For $ICC_{Match}$, only role-specific match performances above 30 minutes are considered, and for $ICC_{Season}$ and $r_{Repeat}$, player-role combinations below 300 minutes of total playing time are excluded. In the Frauen-Bundesliga and World Cup, where the number of matches is more limited, the threshold is reduced to 150 minutes. The role group is added as a covariate to the statistical models behind all meta-metrics to adjust for positional differences.

### 4.3 Summary scores

To provide a categorical summary of the analysed metrics, Validity and Robustness scores are derived by calculating z-scores of all corresponding metrics over the analysed KPIs and competitions, as shown in Equations (7) and (8). These scores are subsequently averaged across the three data sets to yield an overall evaluation of each metric's performance.

$$Validity = \frac{z(r_{MV}) + z(r_{FIFA})}{2} \tag{7}$$

$$Robustness = \frac{z(ICC_{Match}) + z(r_{Repeat}) + z(ICC_{Season})}{3} \tag{8}$$

## 5 Results

After removing role-player combinations with an insufficient number of minutes, 159 performances in the World Cup (60 from central defenders), 236 performances in the Frauen-Bundesliga (47 from central defenders), and 360 performances in the 3. Liga (81 from central defenders) are retained.

Figure 6 exemplarily shows the correlation of the number of interceptions per 90 minutes and valued fault with external ratings for the Frauen-Bundesliga 23/24, illustrating that the classical interceptions metric fails for defenders, i.e. central defenders (CB) and full-backs (FB).

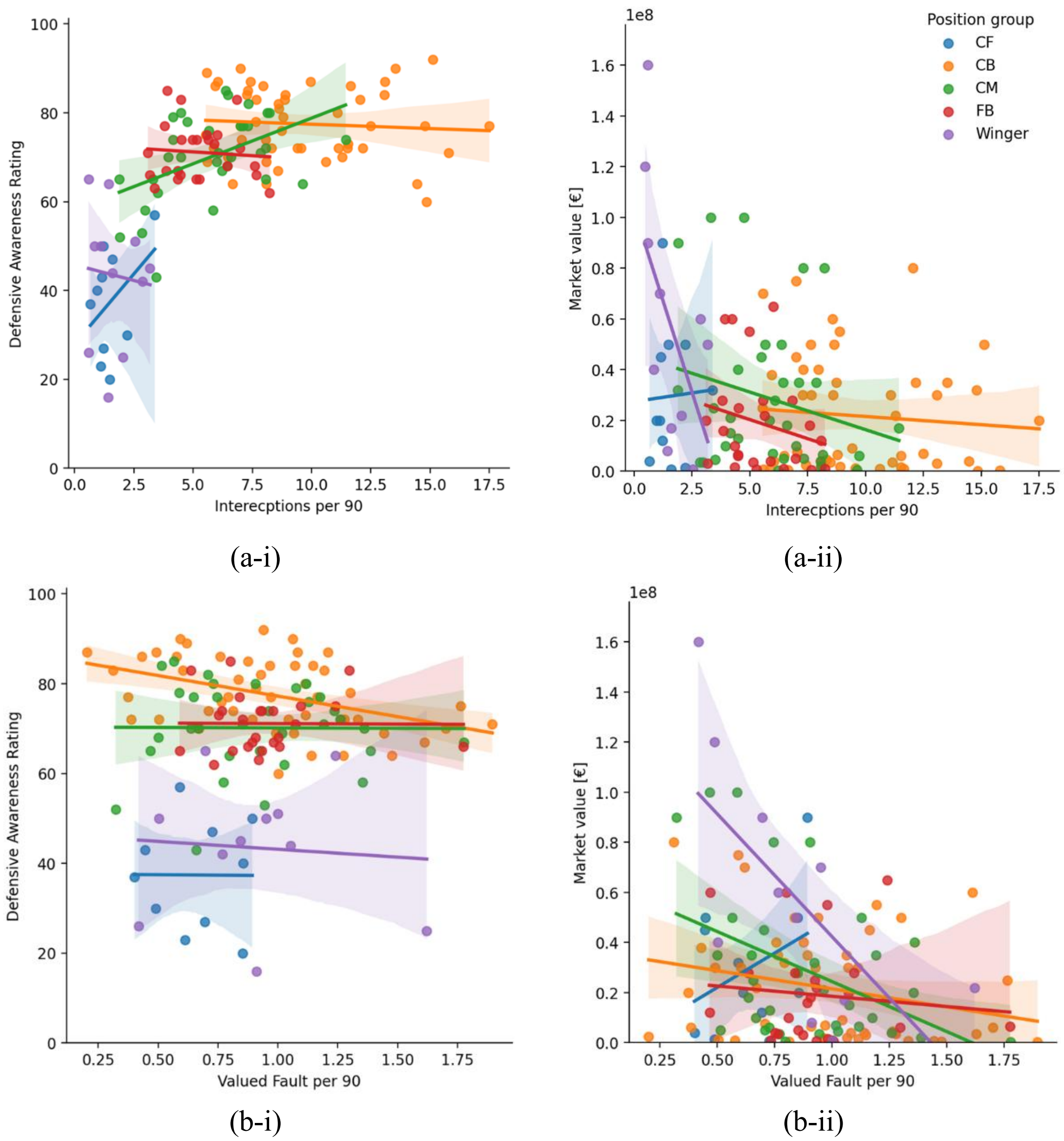

(a-i) (a-ii)

(b-i) (b-ii)

Figure 6. Correlation of the number of (a) interceptions per 90 minutes and (b) valued fault per 90 minutes with the (i) "Defensive Awareness" FIFA rating and (ii) market value.

Figure 7 shows that the strongest metrics with above-average validity and robustness are the fault-related metrics *raw fault per 90 minutes* and *valued fault responsibility per 90 minutes*, as well as the fused metrics *raw involvement*, *raw responsibility* and *valued responsibility*. Also above average score *raw contribution per pass* and *interceptions per pass*. The metric that shows the highest validity but slightly below-average robustness is *valued fault per 90 minutes*.

It is evident from Figure 7 that fault-based metrics perform much better per 90 minutes than per pass while contribution-based metrics perform much better per pass than per 90 minutes. The naïve combined metrics involvement and responsibility consequently perform weakly, while the custom fused versions perform well.

In terms of robustness, raw metrics generally perform better than valued metrics and responsibility-based metrics generally perform better than involvement-based metrics while the effects of both on validity are mixed. Fault generally performs better than contribution with *raw contribution per pass* being the only competitive contribution-based metric. While many contribution-based metrics show strong correlations with external ratings, these correlations are reversed: The higher the contribution, the lower the external rating of the player.

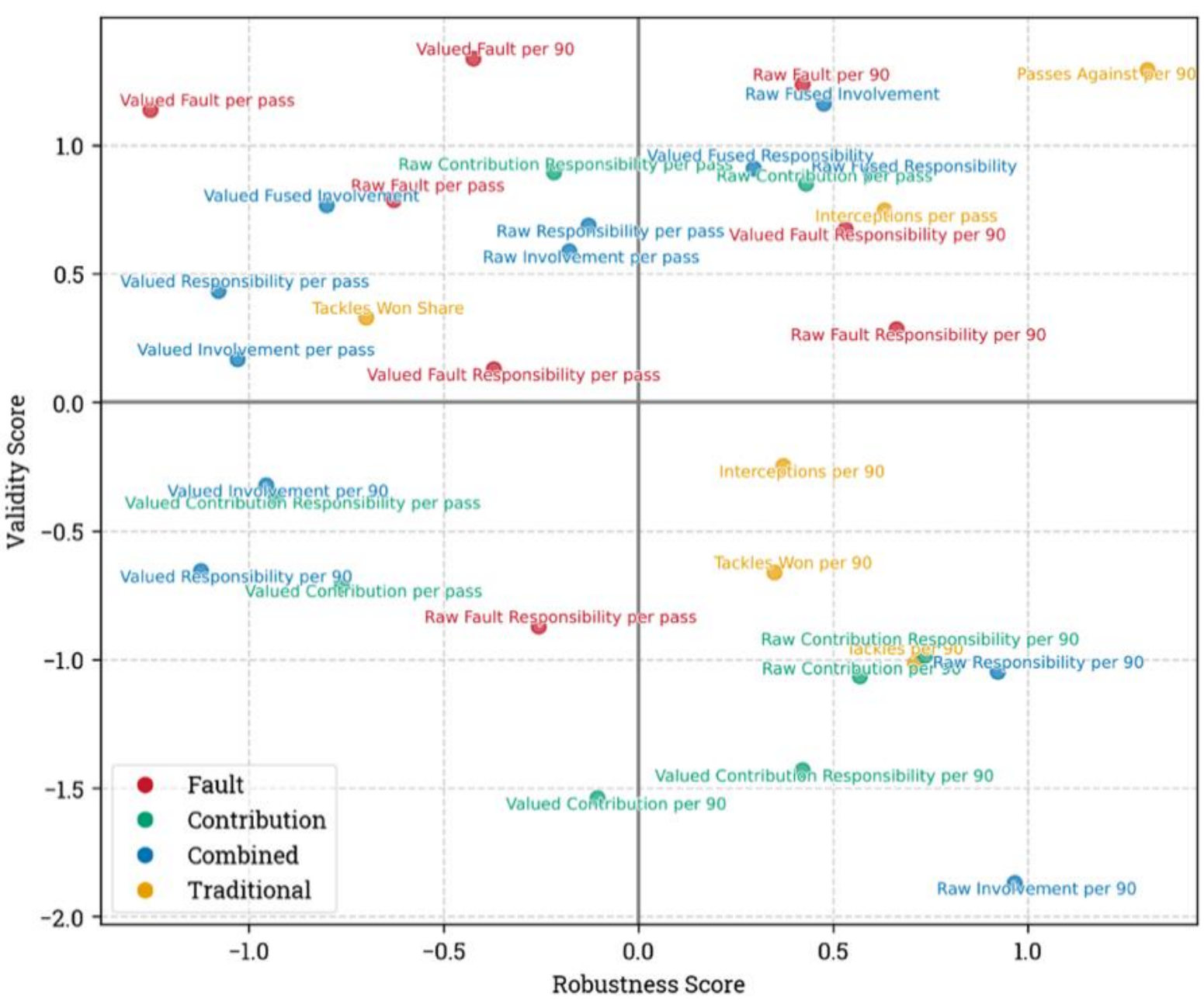

Figure 7. Validity and Robustness scores of all analysed metrics. Higher values on the y-axis indicate stronger correlations with external ratings, and higher values on the x-axis indicate greater stability and discriminatory ability.

Figure 8 compares all metrics across competitions. It is clear that our novel metrics outperform classic metrics in the FIFA Men’s World Cup and Frauen-Bundesliga. In the 3. Liga however, the strongest metrics only marginally outperform the number of interceptions per 90 minutes and are slightly outperformed in terms of robustness.

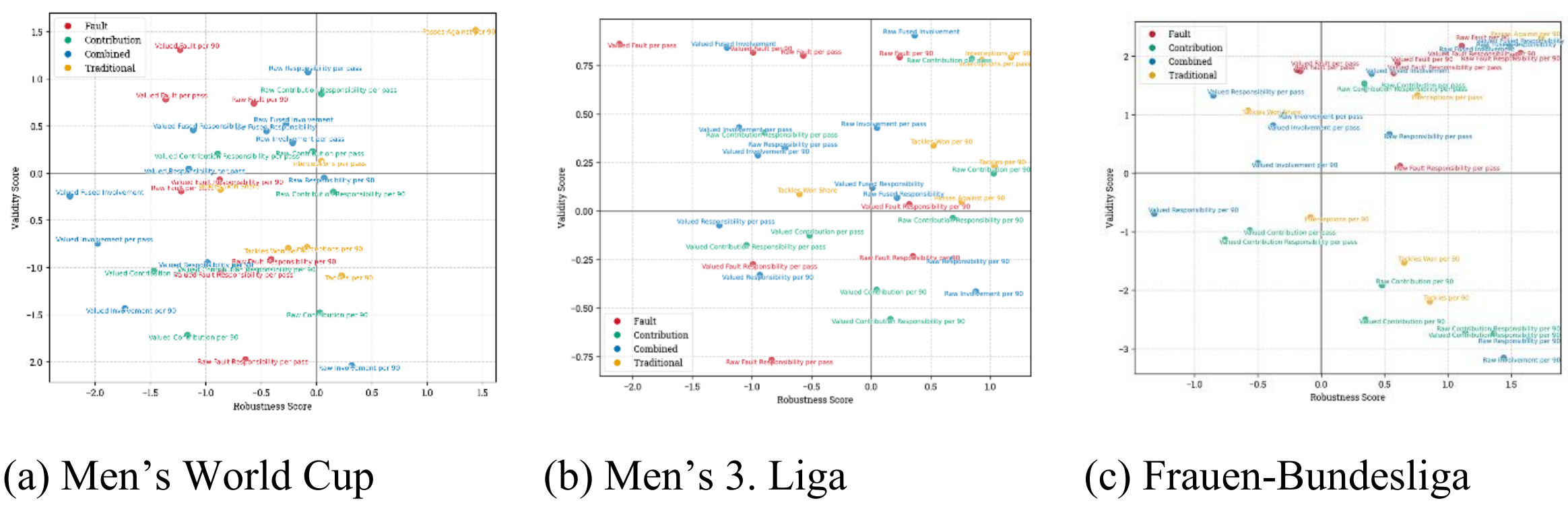

(a) Men's World Cup (b) Men's 3. Liga (c) Frauen-Bundesliga

Figure 8. Comparison of Validity and Robustness scores across competitions.

Figure 9 exemplarily displays the central defenders at the men's World Cup in terms of valued fault and raw contribution, two complementary metrics with the highest validity score. Table 4 and 5 show the top ten central defenders and full backs in the women's Bundesliga according to fused involvement and responsibility.

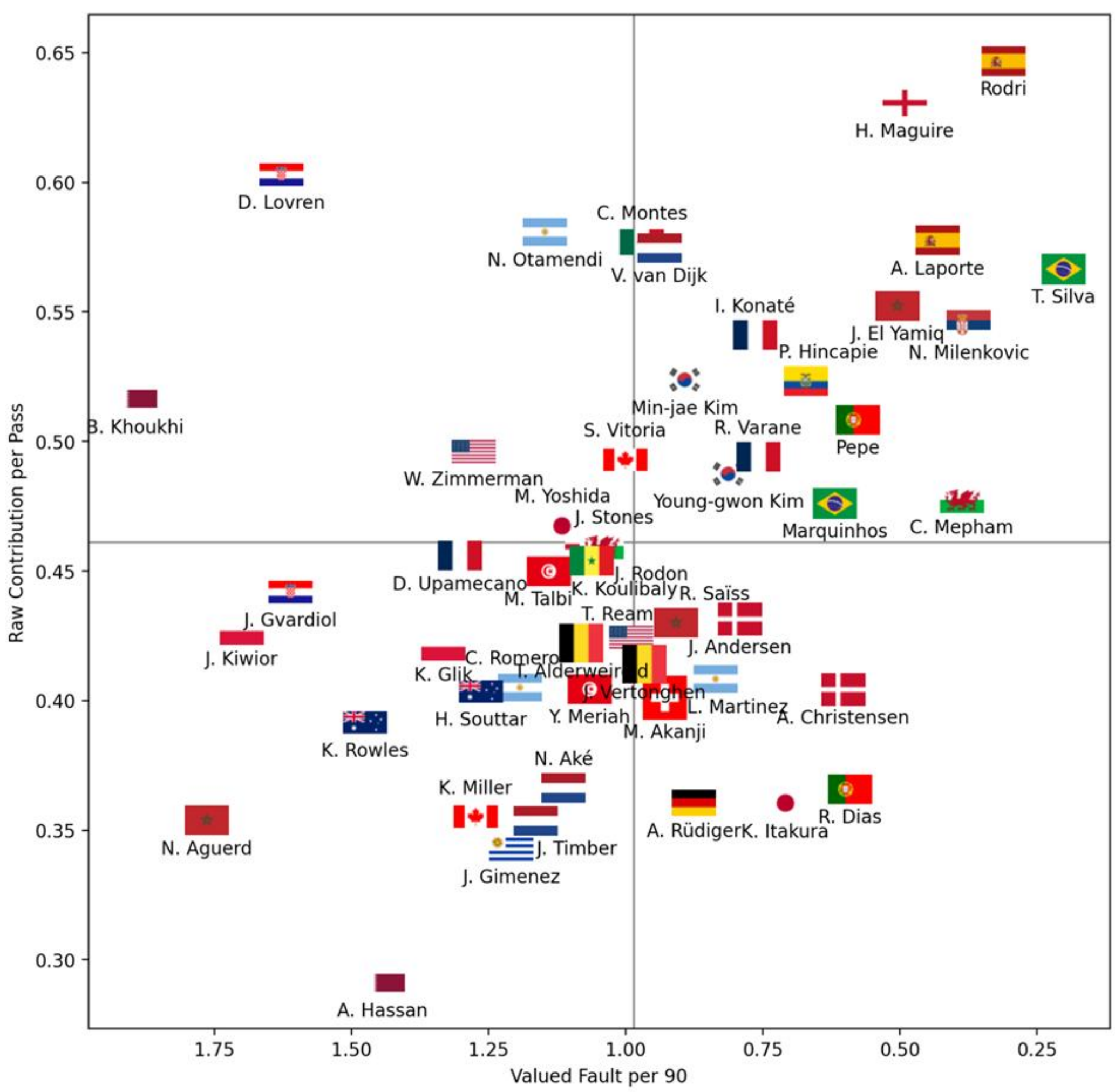

Figure 9. Players plot for valued fault per 90 and raw contribution per pass. Only central

defenders with more than 150 minutes spent in that role are included.

Table 4. Top 10 central defenders in the Frauen-Bundesliga 2023/24 according to fused valued involvement.

| Player | Team | Fused valued involvement | Minutes as CB | Age (2024) | Market value 2024 (€) | Market value rank |
|---|---|---|---|---|---|---|
| Magdalena Eriksson | Bayern München | 1.86 | 537 | 30 | 165,000 | 2 |
| Marina Hegering | VfL Wolfsburg | 1.56 | 615 | 33 | 60,000 | 10 |
| Glódís Viggósdóttir | Bayern München | 1.46 | 950 | 28 | 165,000 | 2 |
| Alina Axtmann | SC Freiburg | 0.28 | 405 | 18 | 25,000 | 32 |
| Michaela Specht | 1899 Hoffenheim | -0.12 | 563 | 26 | 45,000 | 14 |
| Tainara | Bayern München | -0.18 | 181 | 24 | 125,000 | 6 |
| Linda Sembrant | Bayern München | -0.19 | 266 | 36 | 30,000 | 26 |
| Dominique Janssen | VfL Wolfsburg | -0.26 | 594 | 29 | 180,000 | 1 |
| Sara Doorsoun | Eintracht Frankfurt | -0.46 | 956 | 32 | 75,000 | 8 |
| Samantha Steuerwald | SC Freiburg | -0.55 | 516 | 25 | 45,000 | 14 |

Table 5. Top 10 full backs in the Frauen-Bundesliga 2023/24 according to fused valued responsibility.

| Player | Team | Fused valued responsibility | Minutes as FB | Age (2024) | Market value 2024 (€) | Market value rank |
|---|---|---|---|---|---|---|
| Tuva Hansen | Bayern München | 0.073 | 246 | 26 | 75,000 € | 10 |
| Lynn Wilms | VfL Wolfsburg | 0.049 | 562 | 23 | 100,000 € | 7 |
| Nuria Rábano | VfL Wolfsburg | 0.022 | 558 | 24 | 125,000 € | 6 |
| Giulia Gwinn | Bayern München | 0.010 | 717 | 24 | 225,000 € | 1 |
| Lisann Kaut | 1899 Hoffenheim | -0.021 | 190 | 23 | 10,000 € | 48 |
| Katharina Naschenweng | Bayern München | -0.023 | 625 | 26 | 150,000 € | 3 |
| Verena Hanshaw | Eintracht Frankfurt | -0.032 | 779 | 30 | 60,000 € | 14 |
| Joelle Wedemeyer | VfL Wolfsburg | -0.033 | 361 | 27 | 60,000 € | 14 |
| Pia-Sophie Wolter | Eintracht Frankfurt | -0.033 | 859 | 26 | 70,000 € | 12 |
| Felicitas Rauch | VfL Wolfsburg | -0.037 | 210 | 27 | 150,000 € | 3 |

# 6 Discussion

Our results show that involvement- and responsibility-based metrics provide a clear, measurable benefit over the state-of-the-art to capture defensive positioning in terms of validity and robustness. *Raw fault*, *raw involvement* and *valued responsibility* overall outperform the strongest classic metric for defensive positioning, *interceptions per pass*, in terms of validity and robustness across genders and competition levels. *Valued fault* and *raw contribution* show lower robustness yet strong validity. Our metrics also take rich context into account: Pass value, as well as the exact defender location and their role within an automatically recognized tactical setup.

Interestingly, fault-related metrics generally seem to outperform contribution-based metrics. This aligns with the nature of defending, which is fundamentally about preventing mistakes. Defenders can perform on a high level for most of a match, but one mistake is enough to lead to a goal and potentially cost their team points. Positive contributions like winning the ball and exerting pressure are also important but do not carry the same weight as critical failures to prevent danger.

We find two properties of our metrics that regulate them towards higher stability: Adding a location-independent responsibility and removing pass value. Both approaches allow for lower variance on the event level and thus lead to more easily repeatable metrics while retaining the ability to distinguish players. Yet, *valued responsibility* remains strong in both robustness and validity, while *valued fault* achieves the highest validity out of all metrics albeit lower robustness.

We also observe that the effectiveness of metrics differs drastically depending on the aggregation method: Per 90 minutes works better for fault metrics, where the simple accumulation of mistakes indeed likely leads to a worse performance judgement. Per pass however works better for contributions and interceptions where otherwise teams with lower possession would automatically score better due to having more opportunities to defend. Since contribution and fault require different aggregation methods, combining them into an involvement or responsibility score requires a non-trivial fusion of these aggregation methods. Indeed, we observe that the fused aggregation of fault and contribution strongly outperforms naïve responsibility and involvement scores. Of the four fused metrics, only *valued involvement* falls short in terms of robustness as it allows high variation due to its inclusion of both action value and precise player locations.

Our metrics outperform traditional metrics in both men's and women's elite-level competitions. However, in the men's 3. Liga, we find a lower overall level of validity with our metrics and classic metrics achieving similar levels of validity. An explanation for this might be the even skill level in this competition: The best team of the 3. Liga 2023/24, SSV Ulm, scored 2.03 points per match while the weakest one, SC Freiburg II, achieved 0.79. In comparison, Bayern München won the championship in the Frauen-Bundesliga 2023/24 with 2.73 points per match while MSV Duisburg finished last with 0.18 points per game, representing a much larger performance gap in that league compared to the 3. Liga. A large skill gap also exists in the men's World Cup, where the world's elite teams with estimated market values above 1 billion € compete with teams that are valued almost two orders of magnitude lower: in the 2022 edition, England was estimated as the most valuable participant at 1.26 billion Euros while Qatar was the team with the lowest market value, estimated at merely 14.9 million Euros according to "transfermarkt.de". Steeper skill differences between players are easier for a metric to pick up than small differences which might explain the better performance of involvement- and responsibility-based metrics in elite-level competitions. Additionally, an even skill distribution could lead to more tightly contested matches which promote combative playing styles where active defensive contributions become more highly valued than passive ones.

The optimal metric does not just seem to depend on the competition but also on the defender's role: As shown in Figure 7, the validity of fault seems to be slightly higher than the validity of responsibility fault for central defenders. However, if we take the example of Kylian Mbappé who is known to have played a role with little defensive involvement in the World Cup 2022, his raw fault places him as the third best defending winger in the world cup while according to raw fault responsibility, he is placed as the third weakest defending winger. This contrasts with many offensive metrics like pass completion rate and expected threat where the same metric can be used to measure similar skills across roles.

# 7 Limitations

Team strength is an important confounding factor in our validation analysis: In football, ball possession is strongly correlated with general team and player quality: We can observe this from the strong negative correlation of *passes* *against per 90 minutes* with market values. All negative per-90 metrics like fault are therefore inflated to some degree as they automatically punish teams more that spend a lot of time defending and less time in possession. Conversely, per pass aggregation favours players from high possession teams with regards to positive contributions, as they typically play in more aggressive defending schemes that allow them to perform a higher number of effective interceptions and pressures per opponent pass. Since it is not clear whether both effects occur to a similar extent, comparisons between per-90 and per-pass metrics require caution. In addition, the validation against market values and FIFA ratings comes with obvious confounders like age, popularity, and a lack of specificity in measuring defensive off-ball skill. The optimal way to assess the validity of player performance indicators given the lack of a gold standard is an open question in sports analytics.

The techniques presented in this paper attempts to limit complexity by using simple geometric considerations instead of taking advanced variables like pass duration, pressure intensity or body orientation into account. The metrics are designed and tested to yield robust overall evaluations of players and teams rather than detailed insight into specific situations. Future enhancements could build on our baseline by deploying more sophisticated measures of involvement and responsibility, focusing more on individual accountability in specific situations.

Bias could be introduced by additional factors: For example, defensive metrics are typically strongly influenced by the behaviour of the attacking team, which can systematically affect the ratings of players. For example, players who excel defensively might receive more playing time against offensively stronger teams which could deflate their ratings. Future approaches could take this effect into account by comparing defensive metrics to an expected baseline model, such as by relating involvement-based metrics to responsibility-based metrics. The responsibility model might also be biased towards weaker teams and more passive defending styles as these typically allow more passes which are therefore overrepresented in the data. This could be resolved by accounting for additional variables like team identity or pressing style in the model or using resampling techniques.

Foundational modelling components like expected threat and expected receiver can introduce bias, too. It is currently not known how well xT generalizes between genders and competition levels. The domain transfer might therefore introduce a competition-dependent systematic bias. Also, the expected receiver estimate is optimized for computational efficiency, and its accuracy might be surpassed by more advanced approaches. Lastly, while the breadth of the data set used in this study is exceptional, its quality and potential resulting biases are hard to estimate due to the lack of available validation studies.

# 8 Conclusion

This work introduces metrics based on *involvement* and *responsibility* as novel player performance indicators to make the under-appreciated skill of defensive positioning quantifiable. Involvement-based metrics assign fault or contribution for pass-related possession value changes among defenders based on their spatial proximity to the event while responsibility-based metrics represent the “expected involvement” based on player roles within automatically identified tactical formations. Our metrics display higher validity and robustness than traditional defensive metrics based on tackles and interceptions across a uniquely broad data set, spanning genders and competition levels. By including a full season of women’s and lower-level professional men’s football, we demonstrate the applicability of our metrics on players who are underrepresented in the current research landscape.

We find that both fault (involvement and responsibility related to high-valued passes of the opponent) and overall involvement and responsibility ratings are strongly associated with external evaluations of central defenders in elite competitions of both men’s and women’s football. Generally, fault-based metrics show stronger performance than contribution-based metrics while also being a particularly novel aspect of our modelling approach, measuring a failure to prevent dangerous passes. Overall, we find that there is no “one-size-fits-all” solution within the presented involvement-responsibility paradigm, as we find differences in metrics’ optimality between competition levels and player roles. Our work introduces advanced metrics for the intricate problem of assigning individual defensive responsibility for offensive actions, improving our understanding of off-ball defensive ability in sports.

## Statements and Declarations

## Availability of Data and Materials

The World Cup data is publicly available through the data provider: https://www.blog.fc.pff.com/blog/enhanced-2022-world-cup-dataset. The data from the Frauen-Bundesliga and 3. Liga is proprietary and can be requested from *Deutscher Fußball-Bund* (DFB). The code underlying this study is available on Github: https://github.com/jonas-bischofberger/defensive-network.

## Competing interests

The authors declare that they have no competing interests.

## Funding

RM is funded by the China Scholarship Council (No. 202206520005).

## Authors' contributions

RM and JB conceptualized and implemented the work and wrote the manuscript. PB and KA provided access to part of the underlying data and gave feedback on the manuscript. AB supervised the work and gave feedback on the manuscript.

## Acknowledgements

We thank Maike Klemmer for helping us access the data.